\documentclass[prl,twocolumn,showpacs,amsmath,amssymb,superscriptaddress,preprintnumbers]{revtex4}

\pdfoutput=1

\usepackage{graphicx}
\usepackage{dcolumn}
\usepackage{bm}
\usepackage[usenames,dvipsnames]{color}
\usepackage[colorlinks=true, linkcolor=BrickRed, citecolor=Blue, urlcolor=Blue, filecolor=Blue]{hyperref}

\begin{document}

\preprint{Phys.~Rev.~Lett.~\textbf{103}, 211301 (2009)}

\title{Gamma-Rays from Ultracompact Primordial Dark Matter Minihalos}

\author{Pat Scott}
\affiliation{Oskar Klein Centre for Cosmoparticle Physics, AlbaNova, SE-10691 Stockholm, Sweden}
\affiliation{Department of Physics, Stockholm University, AlbaNova, SE-10691 Stockholm, Sweden}

\author{Sofia Sivertsson}
\affiliation{Oskar Klein Centre for Cosmoparticle Physics, AlbaNova, SE-10691 Stockholm, Sweden}
\affiliation{Department of Theoretical Physics, Royal Institute of Technology (KTH), AlbaNova, SE-10691 Stockholm, Sweden}

\date{\today}

\begin{abstract}

Ultracompact minihalos have recently been proposed as a new class of dark matter structure.  These minihalos would be produced by phase transitions in the early Universe or features in the inflaton potential, and constitute non-baryonic massive compact halo objects (MACHOs) today.  We examine the prospect of detecting ultracompact minihalos in gamma-rays if dark matter consists of self-annihilating particles.  We compute present-day fluxes from minihalos produced in the $e^+e^-$ annihilation epoch, and the QCD and electroweak phase transitions in the early Universe.  Even at a distance of 100\,pc, minihalos produced during the $e^+e^-$ epoch should be eminently detectable today, either by the \emph{Fermi} satellite, current Air \v{C}erenkov telescopes, or even in archival \emph{EGRET} data.  Within $\sim$1\,pc, minihalos formed in the QCD phase transition would have similar predicted fluxes to the dwarf spheroidal galaxies targeted by current indirect dark matter searches, so might also be detectable by present or upcoming experiments.

\end{abstract}


\maketitle


The identity of dark matter remains one of the key outstanding problems in physics.  Weakly-interacting massive particles (WIMPs) provide a compelling solution \cite{revs} because their weak-scale masses and cross-sections make for a natural explanation of the observed abundance of dark matter.  As most proposed WIMPs are their own antiparticles, high WIMP densities would also lead to high rates of self-annihilation.  Annihilation products might then provide indirect evidence of the nature of dark matter.  Gamma-rays are particularly attractive in this respect, as they do not suffer the same problems of deflection and attenuation as massive, charged species.

It was proposed \cite{Paczynski86} that dark matter could be massive compact halo objects (MACHOs) of condensed baryons, e.g.~brown dwarfs or faint stars.  These are ruled out as the dominant component of dark matter by the cosmic microwave background (CMB; \cite{wmap5}), Big Bang Nucleosynthesis \cite{IoccoBBN}, and microlensing searches \cite{EROS-OGLE}.  Primordial black holes (PBHs) are an alternative, disfavoured by their energetic evaporation, gravitational influence \cite{Josan09}, and the large primordial density perturbations required for their production ($\delta\gtrsim30$\%).  For comparison, the initial density perturbations from inflation were $\delta\sim10^{-5}$.

Ricotti \& Gould \cite{Ricotti09} proposed a non-baryonic MACHO that avoids these constraints, and presents a promising new target for microlensing searches.  Formation proceeds similarly to PBHs, whereby small-scale density perturbations in the early Universe collapse to a compact body.  A small-scale power spectrum that is the same as observed on large scales \cite{wmap5} provides insufficient power for this to occur.  Perturbations could however be enhanced by features in the inflaton potential, or phase transitions in the early Universe \cite{Schmid97}.  If a perturbation is small, matter will not be sufficiently compressed to form a black hole, leaving only a compact cloud of gas and dark matter.  This mechanism requires density contrasts of just $\delta\gtrsim10^{-3}$ to proceed, so is far more viable than PBH formation.  If such ultracompact minihalos (UCMHs) exist they will be ultra-dense, and excellent targets for indirect detection of WIMPs \cite{note1}.

Here we investigate gamma-ray signals expected from UCMHs containing WIMP dark matter.  We consider UCMHs produced in three phase transitions in the early Universe: electroweak symmetry breaking ($T_\mathrm{EW}\approx200$\,GeV), QCD confinement ($T_\mathrm{QCD}\approx200$\,MeV), and $e^+e^-$ annihilation ($T_{ee}\approx 0.51$\,MeV).  We first discuss the masses, density profiles and primordial abundance of UCMHs, then WIMP models and annihilation channels.  We present predicted fluxes and discuss prospects for detection with satellite missions and Air \v{C}erenkov telescopes (ACTs).  In an appendix, we also give explicit predictions from a supersymmetric framework with a neutralino WIMP.

Following matter-radiation equality, ultracompact minihalos accrete matter by radial infall \cite{Ricotti09} as
\begin{equation}
\label{Mh}
M_\mathrm{h}(z) = \delta m \left(\frac{1 + z_\mathrm{eq}}{1+z}\right),
\end{equation}
where $M_\mathrm{h}(z)$ is the total mass of the UCMH at redshift $z$, and $z_\mathrm{eq}$ is the redshift of matter-radiation equality.  We assume that UCMHs at $z=0$ grew only until $z=10$, because by this time structure formation would have progressed sufficiently far to prevent further accretion \cite{note2}.  The mass contained within a perturbation at equality is
\begin{equation}
\label{deltam}
\delta m = f_\chi\left(\frac{1 + z_\mathrm{eq}}{1+z_\mathrm{X}}\right) M_\mathrm{H}(z_\mathrm{X}),
\end{equation}
where $M_\mathrm{H}(z_\mathrm{X})$ is the horizon mass at the time of phase transition X, and $f_\chi = \Omega_\mathrm{CDM}/\Omega_\mathrm{m} = 0.834$ \cite{wmap5} is the dark matter fraction.  We take $z_\mathrm{eq} + 1 = 2.32\times10^{4}\Omega_\mathrm{m}h^2$ \cite{KolbTurner}, giving $z_\mathrm{eq}= 3160$ with $\Omega_\mathrm{m}h^2 = 0.136$ from the current best fit to the CMB, large-scale structure and Type Ia supernovae \cite{wmap5}.  Eq.~\ref{deltam} arises because only the dark matter fraction of the horizon mass eventually collapses to form a UCMH.  At $z_\mathrm{X}$, this fraction is simply the ratio of the dark matter density to the radiation density, which then evolves linearly with the scalefactor.

During radiation domination, the horizon mass is \cite{ColesLucchin}
\begin{equation}
\label{horizonmassredshift}
M_\mathrm{H}(z) \approx M_\mathrm{H}(z_\mathrm{eq}) \left(\frac{1+z_\mathrm{eq}}{1+z}\right)^2.
\end{equation}
As $T(z) \varpropto g_{*S}(z)^{-\frac{1}{3}}R(z)^{-1} \varpropto g_{*S}(z)^{-\frac{1}{3}}(1+z)$ \cite{KolbTurner}, with $R$ the scalefactor of the Universe and $g_{*S}$ the number of effective entropic degrees of freedom, this becomes
\begin{equation}
\label{horizonmasstemp}
M_\mathrm{H}(T) \approx M_\mathrm{H}(T_\mathrm{eq}) \left(\frac{g_{*S}(T_\mathrm{eq})^\frac{1}{3} T_\mathrm{eq}}{g_{*S}(T)^\frac{1}{3}T}\right)^2.
\end{equation}
The horizon mass, temperature and effective entropic degrees of freedom at equality can be estimated as $M_\mathrm{H}(T_\mathrm{eq}) = 6.5\times 10^{15}\,(\Omega_\mathrm{m}h^2)^{-2} = 3.5 \times 10^{17}$\,M$_\odot$ \cite{Josan09}, $T_\mathrm{eq} = 5.5\,\Omega_\mathrm{m}h^2 = 0.75$\,eV and $g_{*S}(z_\mathrm{eq}) = 3.91$ \cite{KolbTurner}.  At the phase transitions, $g_{*S}(T_\mathrm{EW}) = 107$, $g_{*S}(T_\mathrm{QCD}) \approx 55$ and $g_{*S}(T_\mathrm{ee}) = 10.8$ \cite{KolbTurner}, giving $\delta m_{\{\mathrm{EW,QCD,ee}\}} = \{5.6 \times 10^{-19}, 1.1 \times 10^{-9}, 0.33\}$\,M$_\odot$.

The dark matter density profile in an ultracompact minihalo is \cite{Ricotti09}
\begin{equation}
\label{density}
\rho_\chi(r,z) = \frac{3f_\chi M_\mathrm{h}(z)}{16\pi R_\mathrm{h}(z)^\frac{3}{4}r^\frac{9}{4}},
\end{equation}
in the radial infall approximation, with the maximum extent of the UCMH at redshift $z$
\begin{equation}
\label{Rh}
\left(\frac{R_\mathrm{h}(z)}{\mathrm{pc}}\right) = 0.019\left(\frac{1000}{z+1}\right)\left(\frac{M_\mathrm{h}(z)}{\mathrm{M}_\odot}\right)^\frac{1}{3}.
\end{equation}

The dark matter in an ultracompact minihalo could be further concentrated if baryons collapse and contract the gravitational potential. We calculated the density profile after adiabatic contraction using the method of Blumenthal et al.~\cite{Blumenthal86}. This assumes that $rM(r)$ is conserved at all $r$, where $M(r)$ is the mass within radius $r$, and that orbits of the dissipationless WIMPs do not cross. We assumed that a fraction $F$ of the total halo mass condenses to a constant density baryonic core of radius $r_\mathrm{core}$. We considered $F=10^{-2},10^{-3}$ and $r_\mathrm{core}/R_\mathrm h=5\times10^{-2},10^{-3}$. The effect of the contraction is small for the larger core radius, so we show results only for $r_\mathrm{core}/R_\mathrm h=10^{-3}$. Because the induced contraction at $r$ is given by the increase in the baryonic mass within $r$, the contraction caused by a constant density baryonic core is most pronounced around the core's edge. This is in contrast to the contraction of halos around adiabatically-formed black holes, where the baryons collapse to a central point, steepening the dark matter density profile at all radii.  The dark matter density in the very centre of a halo does not rise significantly in the contraction unless the new baryonic distribution also has a pronounced spike at the very centre.

UCMHs also erode over time as dark matter annihilates away; being ultracompact and ancient, this effect is highly significant.  A simple way to estimate the maximum density $\rho_\mathrm{max}$ at time $t$ in a halo born at $t_\mathrm{i}$ is \cite{Ullio02}
\begin{equation}
\label{rcut}
\rho(r_\mathrm{cut}) \equiv \rho_\mathrm{max} = \frac{m_\chi}{\langle \sigma v \rangle (t - t_\mathrm{i})},
\end{equation}
where $m_\chi$ is the WIMP mass and $\langle \sigma v \rangle$ is the annihilation cross section (multiplied by the collisional velocity and taken in the zero-velocity limit).  We truncate the density profiles at $r=r_\mathrm{cut}$, setting the density within this radius equal to $\rho_\mathrm{max}$.  For UCMHs seen today, $t=13.7$\,Gyr \cite{wmap5}.  For non-contracted UCMHs, $t_\mathrm{i} = t(z_\mathrm{eq}) = 59$\,Myr \cite{Wright06}, because they have existed since the time of equality.  For contracted profiles, $t_\mathrm{i} = t(10) = 0.49$\,Gyr \cite{Wright06}, as they were concentrated at $z=10$.

To estimate the cosmological abundance of UCMHs, one integrates the probability distribution of primordial density perturbations between the UCMH formation threshold ($\delta\sim10^{-3}$) and the PBH threshold ($\delta\sim0.3$).  We approximate the distribution as Gaussian \cite{GreenLiddle}, giving a `bare' relic density (i.e.~ignoring post-equality accretion and disruption) of
\begin{equation}
\label{relicdens}
\Omega(z) = \Omega_\mathrm{CDM}(z) \int^{0.3}_{10^{-3}} \frac{1}{\sqrt{2\pi}\sigma(z_\mathrm{X})} \exp\left(-\frac{\delta^2}{2\sigma(z_\mathrm{X})^2}\right)\mathrm{d}\delta.
\end{equation}
Here $\sigma(z_\mathrm{X})^2$ is the variance of perturbations at the time of the phase transition.  Assuming a scale-independent perturbation spectrum of index $n$, and normalising to the perturbations observed in the CMB, $\sigma$ can be approximated as \cite{GreenLiddle}
\begin{equation}
\label{sigma}
\sigma(z_\mathrm{X}) = 9.5\times10^{-5}\left(M_\mathrm{H}(z_\mathrm{X})/10^{56} g\right)^{(1-n)/4}.
\end{equation}
On CMB scales, $n\sim1$ \cite{wmap5}.  However, the CMB probes only a limited number of modes.  A different power law could plausibly dominate at the small scales relevant to UCMH formation; indeed, many inflationary models give a running spectral index \cite{Josan09}, and phase transitions could produce scale-dependent features in the power spectrum \cite{Schmid97}.  The present limit at the scale of PBH/UCMH formation is $n\lesssim1.25$ \cite{GreenLiddle}.  As they grow by a further factor of 290 (Eq.~\ref{Mh}) between equality and $z=10$, UCMHs formed in the $e^+e^-$ annihilation epoch could account for e.g.~$10^{-3}$ of today's dark matter if $n=1.09$ (assuming they all survive structure formation).  For the QCD and electroweak phase transitions, similar abundances could be obtained for $n=1.06$--$1.07$.

The gamma-ray flux from WIMP annihilation, in a solid angle $\Delta\Omega$ and integrated above energy $E_\mathrm{th}$, is
\begin{eqnarray}
\label{flux}
\Phi(E_\mathrm{th},\Delta\Omega) = & \frac{1}{8\pi m_\chi^2}\sum_f\int^{m_\chi}_{E_\mathrm{th}} \frac{\mathrm{d}N_f}{\mathrm{d}E}\mathrm{d}E \langle\sigma_f v\rangle \nonumber\\
& \times\int_{\Delta\Omega}\int_\mathrm{l.o.s.}\rho^2(\Omega,l)\mathrm{d}l\mathrm{d}\Omega,
\end{eqnarray}
where $\mathrm{d}N_f/\mathrm{d}E$ is the differential photon yield from the $f$th annihilation channel.  The final integral runs over the line of sight to the halo.  For a spherically-symmetric halo appearing as a point source at distance $d$, this is 
\begin{equation}
\label{ptsrc}
\Phi(E_\mathrm{th}) = \frac{1}{2d^2m_\chi^2}\sum_f\int^{m_\chi}_{E_\mathrm{th}} \frac{\mathrm{d}N_f}{\mathrm{d}E}\mathrm{d}E \langle\sigma_f v\rangle \int_0^{R_\mathrm{h}}r^2\rho^2(r)\mathrm{d}r.
\end{equation}
We use $d=100$\,pc as our canonical value, but our results can be rescaled to any~$d$.  With a UCMH mass fraction of $10^{-3}$ and $3.2\times10^{4}$\,M$_\odot$ of dark matter within 100\,pc of Earth (assuming an NFW halo \cite{Battaglia06}), we expect $2\times10^{17}$ electroweak UCMHs, $1\times10^{8}$ QCD UCMHs, or about a $30\%$ chance of finding one $e^+e^-$ UCMH within 100\,pc.  At 100\,pc, all UCMHs are point sources to current experiments.

\begin{figure}
\includegraphics[trim = 20 74 20 100, clip = true, width=0.95\columnwidth]{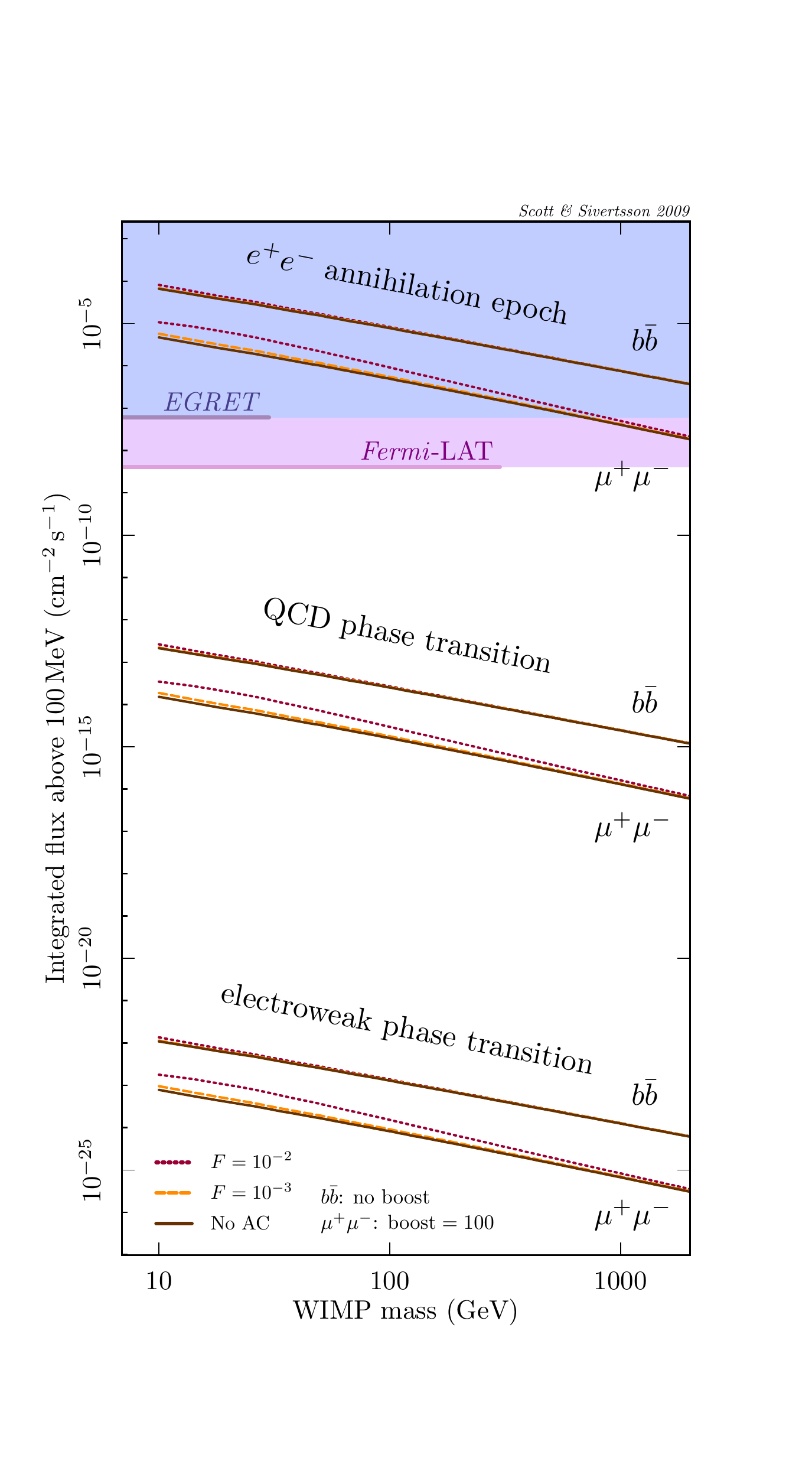}
\caption{Integrated fluxes above 100\,MeV for UCMHs annihilating into either $b\bar{b}$ or $\mu^+\mu^-$ pairs at a distance $d=100$\,pc.  Curves are shown for different phase transitions and degrees of adiabatic contraction.  Adiabatically-contracted UCMHs are assumed to have a fraction $F$ of their mass collapsed into a constant-density baryonic core of radius $10^{-3}R_\mathrm{h}$.  Also shown are approximate $5\sigma$, power-law, high-latitude, point-source sensitivities for 2 weeks of pointed \emph{EGRET} \protect\cite{EGRETsens} and one year of all-sky \emph{Fermi}-LAT \protect\cite{Fermisens} observations.  Solid limits indicate instruments' nominal energy ranges; see also note \protect\onlinecite{finalnote}.}
\label{fig1}
\end{figure}

In Fig.~\ref{fig1} we show gamma-ray fluxes from UCMHs containing WIMPs annihilating into either $b\bar{b}$ or $\mu^+\mu^-$.  We computed these with parton-shower photon yields from \textsf{Pythia 6.4} \cite{Pythia6} in \textsf{DarkSUSY 5.05} \cite{darksusy}.  The $b\bar{b}$ channel is common in supersymmetric models, and the $\mu^+\mu^-$ channel is prominent in models which fit the \emph{PAMELA} and \emph{Fermi} electron excesses \cite{Pamelapositron,Fermielectron}.  For the $b\bar{b}$ channel we use the canonical cross-section $\langle\sigma v\rangle=3\times10^{-26}$\,cm$^3$\,s$^{-1}$ implied by the relic density.  For $\mu^+\mu^-$ we apply a boost factor of 100, corresponding to the minimum Sommerfeld enhancement necessary to explain the electron data in many models.  If a UCMH were situated sufficiently nearby however, its compactness might provide the required boost factor without needing any Sommerfeld enhancement.

\begin{figure}
\includegraphics[trim = 20 74 20 100, clip = true, width=0.92\columnwidth]{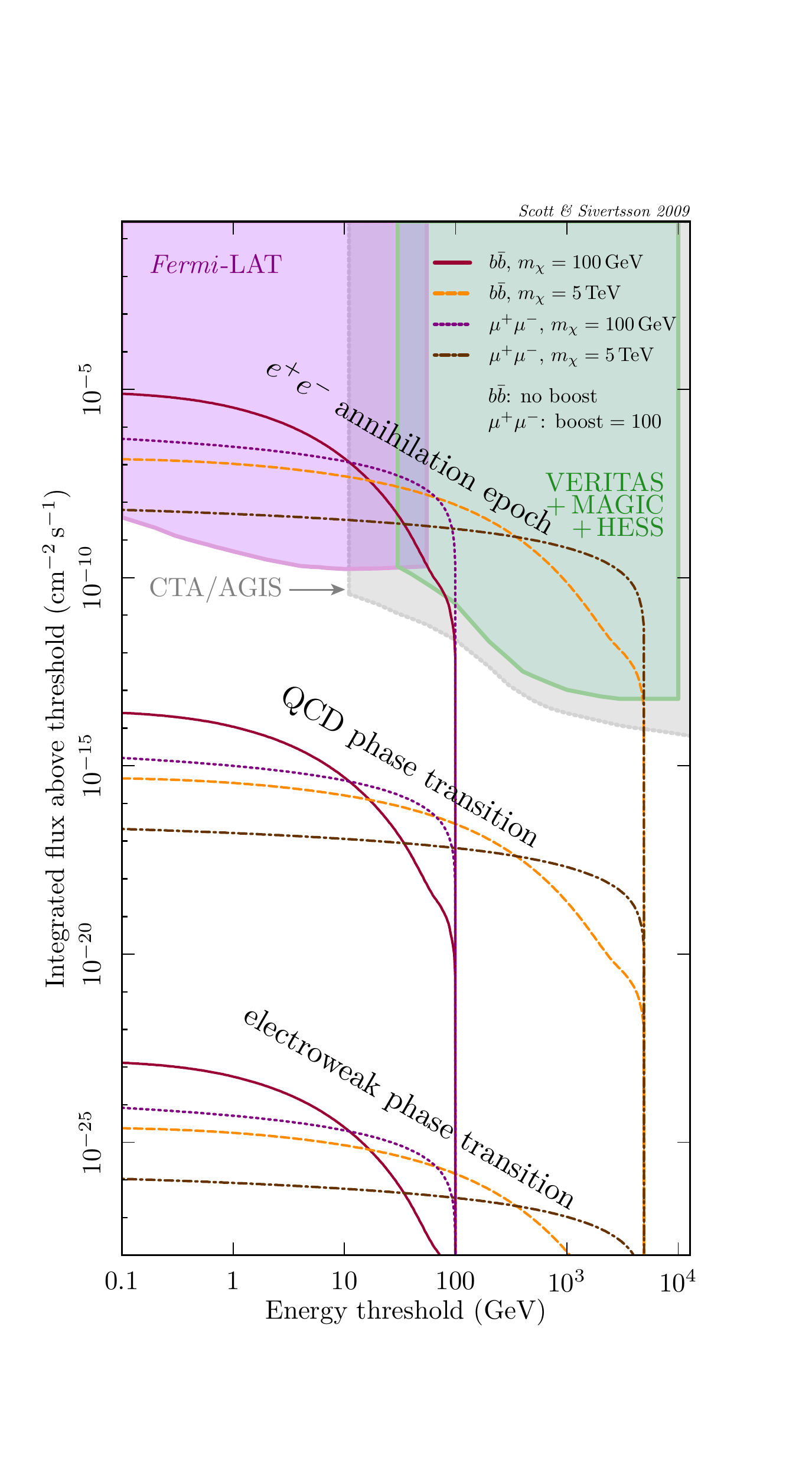}
\caption{Fluxes from uncontracted UCMHs at $d=100$\,pc, as a function of the energy threshold of the observing experiment.  Shaded areas show the regions accessible after a 1 year survey by the \emph{Fermi}-LAT \protect\cite{Fermisens}, and 50\,hr of observation by existing and planned Air \v{C}erenkov Telescopes \protect\cite{CTAricap}.  See also note \protect\onlinecite{finalnote}.}
\label{fig2}
\end{figure}

Despite the increased density, adiabatic contraction does not greatly increase the gamma-ray flux.  This is because the flux profile is dominated by the central region, which is not strongly contracted.  The Sommerfeld enhancement we used for the $\mu^+\mu^-$ channel increases $r_\mathrm{cut}$, making the flux profile less concentrated at the centre and therefore more responsive to increases in density near $r_\mathrm{core}$.  If $r_\mathrm{core} \ll r_\mathrm{cut}$ or $r_\mathrm{core} \gg r_\mathrm{cut}$, this effect is absent.

In Fig.~\ref{fig1}, we show representative point source sensitivities of \emph{EGRET} \cite{EGRETsens} and \emph{Fermi} \cite{Fermisens} above 100\,MeV.  Fig.~\ref{fig2} gives the expected fluxes as a function of threshold energy, allowing for a direct comparison with the sensitivities of current and upcoming ACTs \cite{CTAricap,finalnote}.  

UCMHs formed in the $e^+e^-$ annihilation epoch should be observable by either \emph{Fermi}, MAGIC or HESS, depending upon the WIMP mass.  They could have already been seen by \emph{EGRET} in some cases, effectively ruling out the $b\bar{b}$ channel up to multi-TeV masses.  Given their radial flux profiles, UCMHs from the $e^+e^-$ epoch as close as $d\sim1$\,pc would even appear as extended sources to \emph{Fermi}.  The non-discovery to date of a point source with the spectral characteristics of annihilating dark matter suggests that the amplitude of perturbations generated by $e^+e^-$ annihilation in the early Universe was likely $\delta<10^{-3}$.  A dedicated analysis of the \emph{EGRET} and \emph{Fermi} catalogues (particularly unidentified sources) is required for this statement to be made more definite.  Such a study might even reveal some UCMH candidates.  Limits from ACTs are more difficult to obtain, as UCMHs could have simply been missed by observing the wrong parts of the sky.  On the other hand, if microlensing searches detect a UCMH from the $e^+e^-$ transition, it can potentially be followed up by ACTs.

UCMHs from the QCD phase transition are not yet visible at $d=100$\,pc, but at $d=1$\,pc their predicted fluxes would be comparable to those of dwarf galaxies [e.g. \onlinecite{Martinez09}].  If their abundance and the distance of the nearest example from Earth were favourable, they might be seen by \emph{Fermi} or future instruments like the \v{C}erenkov Telescope Array (CTA).  UCMHs from the electroweak phase transition will not be detectable soon unless some lie within the Solar System; in any case, light UCMHs would face formation problems from kinetic coupling and free-streaming of dark matter.

These results have important implications.  Because of Eq.~\ref{rcut}, the microlensing profiles of UCMHs containing WIMPs could differ from those of Ref.~\onlinecite{Ricotti09}.  The additional annihilation products generated by UCMHs early in their lives could have an impact upon the ionisation history of the Universe, and photons from the extra annihilation might modify the extragalactic gamma-ray background.  If models explaining the \emph{Fermi} and \emph{PAMELA} electron excesses are accurate, UCMHs would also inject more electrons into the intergalactic medium and increase inverse Compton scattering of the CMB at all wavelengths.

\begin{acknowledgments}
We thank Torsten Bringmann, Joakim Edsj\"o, Joachim Ripken, Dave Thomson and the anonymous referees for helpful comments, and the Swedish Research Council for funding support.
\end{acknowledgments}

\bibliography{DMbiblio,SUSYbiblio}

\begin{thebibliography}{26}
\bibitem{note1}{Ricotti \& Gould \cite {Ricotti09} also discuss ultracompact minihalos containing PBHs, but these require similar amplitude density perturbations as PBHs, so are less appealing.}
\bibitem{note2}{Using e.g. $z=30$ instead has little impact on results.}
\end{thebibliography}

\appendix

\section{Appendix: Flux predictions in the CMSSM}

\begin{figure}[hb]
\includegraphics[trim = 40 204 20 160, clip = true, width=0.9\columnwidth]{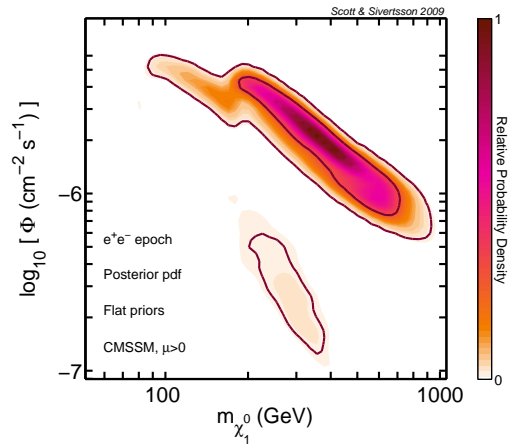}
\caption{Expected fluxes at $d=100$\,pc in the context of the CMSSM, for UCMHs formed in the $e^+e^-$ annihilation epoch, integrated above 100\,MeV.  Contours indicate 1 and $2\sigma$ confidence intervals.  Fits included a range of experimental data, and required that the neutralino is the only component of dark matter.  Predictions from the QCD and electroweak transitions look similar, but are $\sim$8.5 and $\sim$17 orders smaller, respectively.}
\label{fig3}
\end{figure}

Fig.~\ref{fig3} shows neutralino annihilation fluxes predicted in the Constrained Minimal Supersymmetric Standard Model (CMSSM), for UCMHs formed in the $e^+e^-$ epoch.  We performed a global CMSSM fit using \textsf{SuperBayeS} \cite{Trotta08}, including CMB constraints on the relic density, accelerator searches for sparticles and the Higgs boson, the muon $g-2$, the $\bar{B}_s - B_s$ mass difference, and limits on rare \mbox{$B$-decays}.  Details can be found in Ref.~\onlinecite{Trotta08}.  Fluxes show a familiar band of high probability from points in the focus point region, due to clustering around the canonical annihilation cross-section compatible with the relic density.  A lower-probability region is also seen, corresponding to models where stau co-annihilation is significant.  The entirety of the allowed CMSSM parameter space should be accessible by current instruments if UCMHs were formed in the $e^+e^-$ epoch.  Predictions for UCMHs arising in the QCD and electroweak transitions look similar, but are shifted to lower fluxes as in Figs.~\ref{fig1} and \ref{fig2}.

\end{document}